\documentclass[prl,twocolumn,amsmath,amssymb,superscriptaddress]{revtex4}

\usepackage{graphicx}
\usepackage{mathrsfs}
\usepackage[colorlinks]{hyperref}
\usepackage{braket}
\usepackage{subfigure}

\begin{document}

%
% Title
% 

\title{Photon Added Coherent States: Nondeterministic, Noiseless Amplification in Quantum Metrology}

%
% Authors
%

\author{Bryan T. Gard}
\affiliation{Hearne Institute for Theoretical Physics and Department of Physics \& Astronomy, Louisiana State University, Baton Rouge, LA 70803, USA}
\email{bgard1@lsu.edu}
\author{Dong Li}
\affiliation{Hearne Institute for Theoretical Physics and Department of Physics \& Astronomy, Louisiana State University, Baton Rouge, LA 70803, USA}
\affiliation{Quantum Institute for Light and Atoms, Department of Physics,
East China Normal University, Shanghai 200062, P. R. China}
\author{Chenglong You}
\affiliation{Hearne Institute for Theoretical Physics and Department of Physics \& Astronomy, Louisiana State University, Baton Rouge, LA 70803, USA}
\author{Kaushik P. Seshadreesan}
\affiliation{Hearne Institute for Theoretical Physics and Department of Physics \& Astronomy, Louisiana State University, Baton Rouge, LA 70803, USA}
\author{Richard Birrittella}
\affiliation{Department of Physics and Astronomy, Lehman College, The City University of New York, Bronx, New York 10468-1589, USA}
\author{Jerome Luine}
\affiliation{Northrop Grumman Aerospace Research Laboratories, Redondo Beach, California 90278, USA}
\author{Seyed Mohammad Hashemi Rafsanjani}
\affiliation{Institute of Optics, University of Rochester, Rochester, New York 14627}
\author{Mohammad Mirhosseini}
\affiliation{Institute of Optics, University of Rochester, Rochester, New York 14627}
\author{Omar S. Maga\~{n}a-Loaiza}
\affiliation{Institute of Optics, University of Rochester, Rochester, New York 14627}
\author{Benjamin E. Koltenbah}
\affiliation{Boeing Research \& Technology, Seattle, WA 98124, USA}
\author{Claudio G. Parazzoli}
\affiliation{Boeing Research \& Technology, Seattle, WA 98124, USA}
\author{Barbara A. Capron}
\affiliation{Boeing Research \& Technology, Seattle, WA 98124, USA}
\author{Robert W. Boyd}
\affiliation{Institute of Optics, University of Rochester, Rochester, New York 14627}
\author{Christopher C. Gerry}
\affiliation{Department of Physics and Astronomy, Lehman College, The City University of New York, Bronx, New York 10468-1589, USA}
\author{Hwang Lee}
\affiliation{Hearne Institute for Theoretical Physics and Department of Physics \& Astronomy, Louisiana State University, Baton Rouge, LA 70803, USA}
\author{Jonathan P. Dowling}
\affiliation{Hearne Institute for Theoretical Physics and Department of Physics \& Astronomy, Louisiana State University, Baton Rouge, LA 70803, USA}

%
% Abstract
%

\begin{abstract}
Probabilistic amplification through photon addition, at the output of an Mach-Zehnder interferometer is discussed for a coherent input state. When a metric of signal to noise ratio is considered,  nondeterministic, noiseless amplification of a coherent state shows improvement over a standard coherent state, for the general addition of $m$ photons. The efficiency of realizable implementation of photon addition is also considered and shows how the collected statistics of a post selected state, depend on this efficiency. We also consider the effects of photon loss and inefficient detectors.
\end{abstract}

\maketitle

Here we discuss the use of photon addition and subtraction as a probabilistic amplifier and it's effects on coherent light. The mathematical description of adding a photon to coherent light was first proposed by Agarwal and Tara \cite{Agar1, Agar2}. The process of adding photons is a form of noiseless amplification and can be used to enhance a general signal with no added noise, but with the requirement that it does so nondeterministically \cite{Pandey2013}. Unlike many past discussions of this implementation \cite{Gerry2012a, Gerry2014a}, we consider the case of photon addition at the \emph{output} of a Mach-Zehnder Interferometer (MZI) \cite{MZI1}.  Since we are using an MZI model, we are then in the realm of metrology and can therefore use many previously developed techniques from this field. The reasoning behind using the probabilistic amplification operation at the output is simply a model of the limit of control over a specified system. In the case of an externally measured source, meaning a source one has no direct control over, deterministic amplification proves useless for a metric of Signal to Noise Ratio (SNR), since a deterministic amplifier always amplifies signal along with its noise, leaving SNR invariant at best. More concretely, this restriction means any modification to a standard MZI must be done after the phase shifter $\phi$. With this restriction in mind, the question then remains, since a deterministic amplifier does not provide any benefit, is there any hope for a probabilistic amplification process? Another possible benefit of this modification at the output, rather than the input, is that, any generation of a quantum state is performed ``locally, at the detector" and therefore much less exposed to decoherence from the environment. We will show that the SNR may be nondeterministically increased, through the use of photon addition, the effect of the post-selection requirement on simulated data and the behavior of this scheme under photon loss and inefficient detectors.

Recent discussions by Caves \cite{Caves2014} show the use of post-selection schemes and their place in quantum metrology protocols. Indeed, post selection schemes alone do not allow for increased phase information, but this result does not invalidate the usefulness of post selection schemes in metrology, when other metrics are considered.
 
With the goal of describing this scheme in a full model, we begin with a mathematical description of our input states. In order to fully characterize the propagation of light into our MZI, we choose to use the description in terms of propagation of phase space operators and Wigner functions. Following Fig. \ref{fig:MZI}, one particular choice, is a two mode input Wigner function that is a simple product of a coherent and vacuum state,
\begin{equation}
W_{in}({\textbf{X}})=\frac{1}{\pi^2}\textrm{e}^{-(x_1-\sqrt{2}\alpha \cos\theta)^2-(p_1-\sqrt{2}\alpha\sin\theta)^2-x_2^2-p_2^2},
\end{equation}

where ${\textbf{X}}$ encompasses all phase space quadrature mode labels, $\bar{n}=|\alpha|^2$ and $\theta$ is the phase of the coherent state. The propagation of this two mode Wigner function through the various linear optical elements of the MZI is a straightforward task \cite{Kaushik1}. It is important to note that once we have a form of the output state, prior to our probabilistic amplification, all operations are Gaussian preserving, but the use of photon addition or subtraction breaks this preservation (due to its projective measurements). Another important task is how exactly to model the use of photon subtraction (addition). Ideally these can be modeled with annihilation (creation) operators \cite{Cerf2005,Jeong2014,Jeong2016,Parigi2007}, but these assume a non-physical, ideal case (this treatment is only accurate if one restricts to efficiencies $\lesssim 5\%$), and while useful for some treatments, \emph{cannot} be used when accurately accounting for the efficiency of post-selection. This distinction is important when considering phase measurement (phase sensitivity) as ones metric. Recently, this topic has been discussed on how to account for post selection in information terms \cite{David2015}. It is crucial to note that use of post-selection, in any form, requires careful consideration when one quantifies the usefulness of a proposed scheme. This directly affects the efficiency of photon addition and subtraction schemes, since all known realizations of these processes rely on post selection. Specifically, if one is interested in Fisher Information (and therefore phase sensitivity) for a probabilistic process, one calculates $F(\phi)=P_s F_s(\phi)+P_f F_f(\phi)$, where ``$s$'' (``$f$'') subscripts denote success (failure) of the probabilistic process. This process separates the outcomes into complementary cases, in principle however, one can imagine collecting statistics from \textit{both} outcomes. Instead of the mentioned mathematical representation, we choose to model these operations by their experimental implementations. Of course, there exist different ways to implement addition or subtraction of a photon, but here, we choose realistic implementations that are readily available. For photon subtraction, this means use of a variable transmissivity beam splitter. For photon addition, a parametric down conversion heralded process, as well as Fock states incident on a variable transmissivity beam splitter. 

Since our input state here is a coherent state and it has the property that it is an eigenvector of the annihilation operator, $\hat{a} |\alpha \rangle=\alpha |\alpha \rangle$, photon subtraction proves useless for this case as it merely attenuates the coherent state but does not alter the state itself.

\begin{figure}[!htb]
\includegraphics[width=0.7\columnwidth]{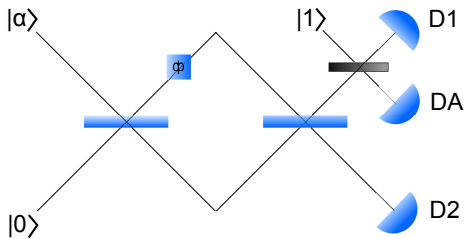}
\caption{A standard Mach-Zehnder Inteferometer where $\phi$ encompasses a phase difference between the two arms, as well as a control phase (not shown). Photon addition is modeled in the top output arm by a variable tranmissivity (T) beam splitter (black) where other beams splitters are standard 50/50. Post selection is conditioned on the successful detection of no photons, registered at detector DA. Note that this depicts single photon addition, but one can easily consider the same model for $m$ photon additions, replacing the single photon state with an $m$ photon state.} \label{fig:MZI}
\end{figure}

\begin{figure}[!htb]
\includegraphics[width=0.35\columnwidth]{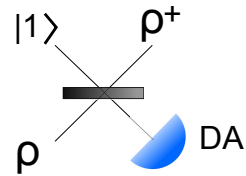}
\caption{Simplistic version of photon addition using a variable transmissivity beam splitter with a single photon Fock state and arbitrary state as input. One output mode is monitored with a detector, conditioned on receiving no photons. Under this condition, the remaining output mode is now known to be photon added. Scales to $m$ photon additions by replacing the single photon state with a $m$ Fock state.} \label{fig:padd}
\end{figure}

Photon addition can be modeled in various ways, for single photon addition, the combination of a single photon Fock state onto a variable transmissivity beam splitter along with the state to be photon added. Shown in Fig. \ref{fig:padd}, one output mode of the beam splitter is monitored to herald when no photons are present, confirming that all light exits the other output mode, showing that the mode is now photon added. Photon addition can also be accomplished with Parametric Down Conversion (PDC) \cite{Bellini2009} where vacuum and the state to be photon added enter a down conversion crystal. Here one output mode (idler mode) is measured for a single photon and when confirmed, the signal mode is known to be photon added, as under ``weak seeding", spontaneous down conversion is the dominant process, which always emits a pair a photons into each output mode. In the beam splitter treatment the coupling parameter is controlled by the transmissivity (or reflectivity) of the variable beam splitter. For the down conversion case, the coupling parameter is controlled by the gain, $G\geq 1$, of the down conversion process. This gain factor is related to the squeezing strength by $G=\textrm{cosh}^2(r)$, where $r$ is the squeezing parameter.

\begin{figure}[!htb]
\includegraphics[width=\columnwidth]{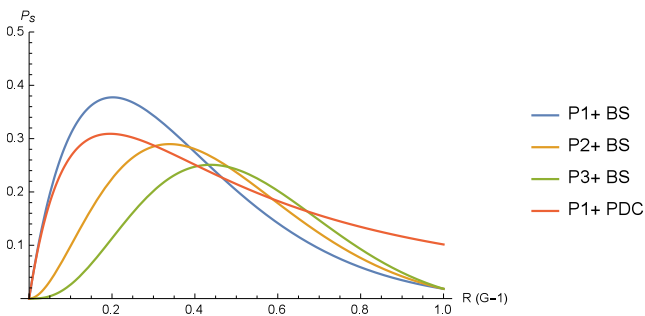}
\caption{Probability of successfully heralding photon addition, as a function of the reflectivity, to a coherent with a variable beam splitter for one, two and three photon additions. Also for comparison is the probability of adding a single photon by a PDC process as a function of gain. Notice the similar statistics between single photon addition using a beam splitter and PDC, but also note that in general $0 \leq R \leq 1$, but $1 \leq G \leq \infty$.} \label{fig:probscoh}
\end{figure}

In the case of photon addition with a variable beam splitter, post selecting on the events that the detector, DA, receives no photons, amounts to forming a projective measurement on the state \cite{Kim2008} in the form of

\begin{equation}
\begin{split}
&W_{\textrm{out}}(x_1,p_1,x_2,p_2)=\int W_{\textrm{tot}}(\textbf{X})\times\\
&(2\textrm{e}^{-x_3^2-p_3^2}) dx_3 dp_3,
\end{split}
\end{equation}

where $\textbf{X}$ represents all phase space variables of the total state $(W_{\textrm{tot}})$ after the variable beam splitter and mode 3 is the mode performing the projective measurement at detector DA. This projects the state into the subspace when detector DA receives only vacuum. This resulting state requires proper normalization and this normalization is also the probability that the detector DA succeeds \cite{Kim2008}. This probability is given by

\begin{equation}
P_1=-(T-1)(1+T|\alpha|^2)\textrm{e}^{|\alpha|^2(T-1)},
\label{eq:padd}
\end{equation}
where $T$ is the transmissivity of the variable beam splitter and we have set $\phi=\theta=0$ so that the maximum amount of coherent light enters the mode to be added. Unless otherwise stated, $\phi=0$ is assumed for the remainder the paper. If one wishes to model the addition of more than one photon, the advantage of this simplistic model becomes apparent. If we wish to add multiple photons to the coherent state, we simply send in higher Fock states into the variable beam splitter but do not need to alter our post selection scheme. Indeed, we find that for $m$ photon additions to a coherent state under the beam splitter model, the probability of generating a $m$ photon added coherent state goes as,

\begin{equation}
P_m=(1-T)^m\textrm{e}^{|\alpha|^2(T-1)} L_m(-T|\alpha|^2 ),
\label{eq:paddm}
\end{equation}
where $m$ is the number of added photons and $L_m(x)$ is the Laguerre polynomial of order $m$. These probabilities, for up to $m=3$, are shown in Fig. \ref{fig:probscoh}. This figure also shows the comparison for single photon addition with the BS and PDC models.

In the case of using a beam splitter for photon addition, several hurdles would need to be overcome. Generation of single photon states (or higher order photon states) on demand is not a trivial task but has made significant progress in recent years \cite{Asta2015} and in general would enter into the efficiency of this scheme, but in this discussion we ignore this portion of the efficiency and only consider the efficiency of the process shown in Fig. \ref{fig:padd}. Also of concern is the use of monitoring an output mode for vacuum, confirming when an addition has taken place, which is severely dependent on the efficiency of the chosen detector (large dark counts would hurt this process significantly) as well as restricting the state to be added to low photon numbers. Higher-order photon addition can be accomplished in much the same way as with the single photon state, only the input state needs to be altered (post selection remains the same).

In the case of using PDC as a source for photon addition, other concerns are worth noting. Use of a non-linear crystal itself, while somewhat standard practice, is itself far more involved than a much simpler passive linear element such as a variable transmissivity beam splitter. Beyond this, much like in the beam splitter case, we are restricted to low photon number sources due to wanting SPDC as the dominant process. Higher order photon addition is accomplished by changing the conditional measurement in the idler mode from receiving a single photon, to receiving $m$ photons, confirming that the signal mode has been $m$ photon added. So the scalability here requires the use of photon number resolving detectors, which are becoming much more readily available recently \cite{TES1,Clem2010a}. We note that the performance of the beam splitter and PDC model are very similar and therefore focus on the simpler beam splitter model.

Photon addition and subtraction have many properties that appear, at first glance, unintuitive, but nevertheless, have been demonstrated in experiment \cite{Zavatta2004a,Zavatta2005a,Bellini2009,Our2006a,Wenger2004a, Nielsen2006a, Wakui2007a}. Of these, perhaps the initial surprising aspect is that, under the process of photon subtraction, the average photon number actually increases (except in the case of a coherent state, where the state is unchanged). In this case, this effect is attributed to the fact that the post selection stage projects the remaining state into a state with higher average photon number, since partial knowledge of the state, specifically that it likely does not contain a significant vacuum contribution, is measured.

\begin{figure}[!htb]
\includegraphics[width=\columnwidth]{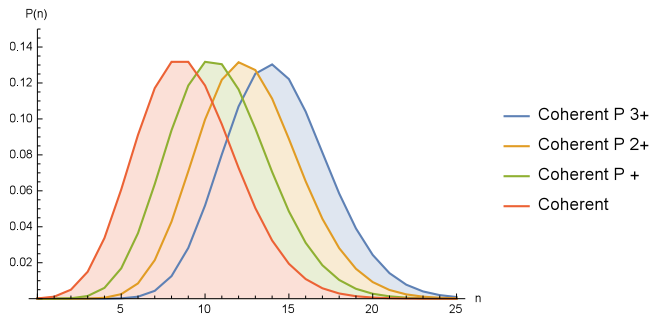}
\caption{Photon number distributions for a normal coherent state along with a photon added coherent state for values $m=1,2,3$. A clear shift in photon number distribution can be see as more photons are added. Tranmissivity of the beam splitter is set to $T=1$. We note (but not shown here) that single photon addition to a coherent state through the PDC model gives a near identical photon number distribution for similar interaction parameters as compared to this BS model.} \label{fig:paddcoh}
\end{figure}

\begin{figure}[!htb]
\centering
\begin{subfigure}
{\includegraphics[width=\columnwidth]{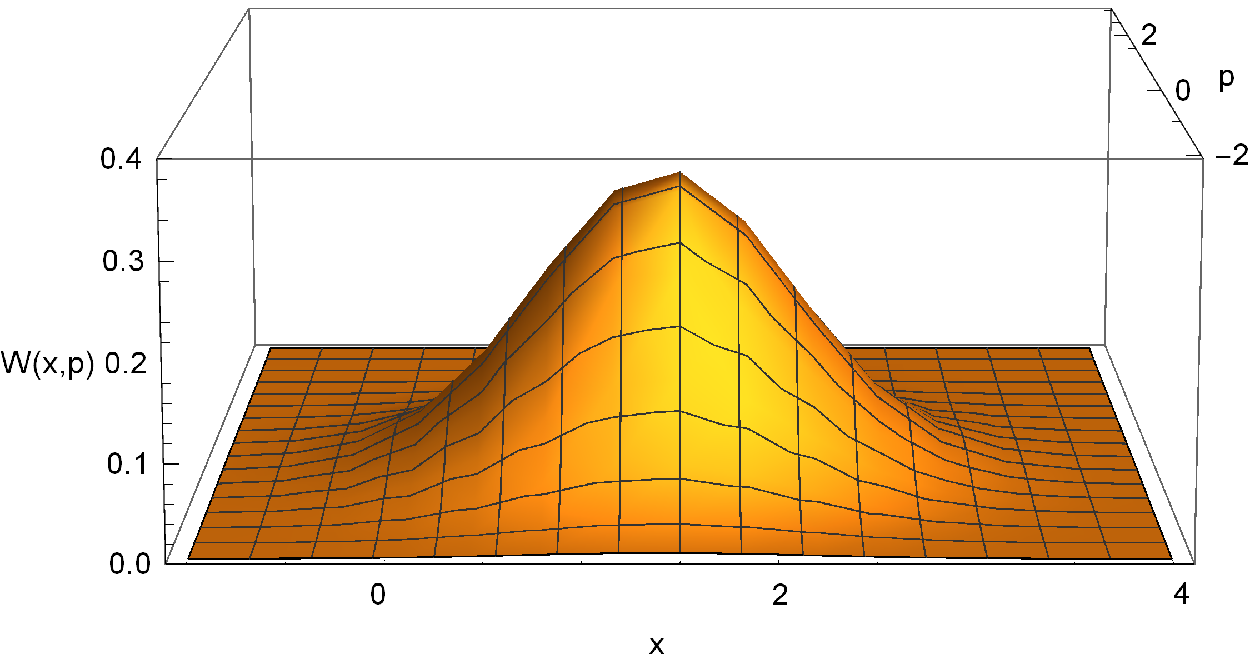}}
\end{subfigure}
~
\begin{subfigure}
{\includegraphics[width=\columnwidth]{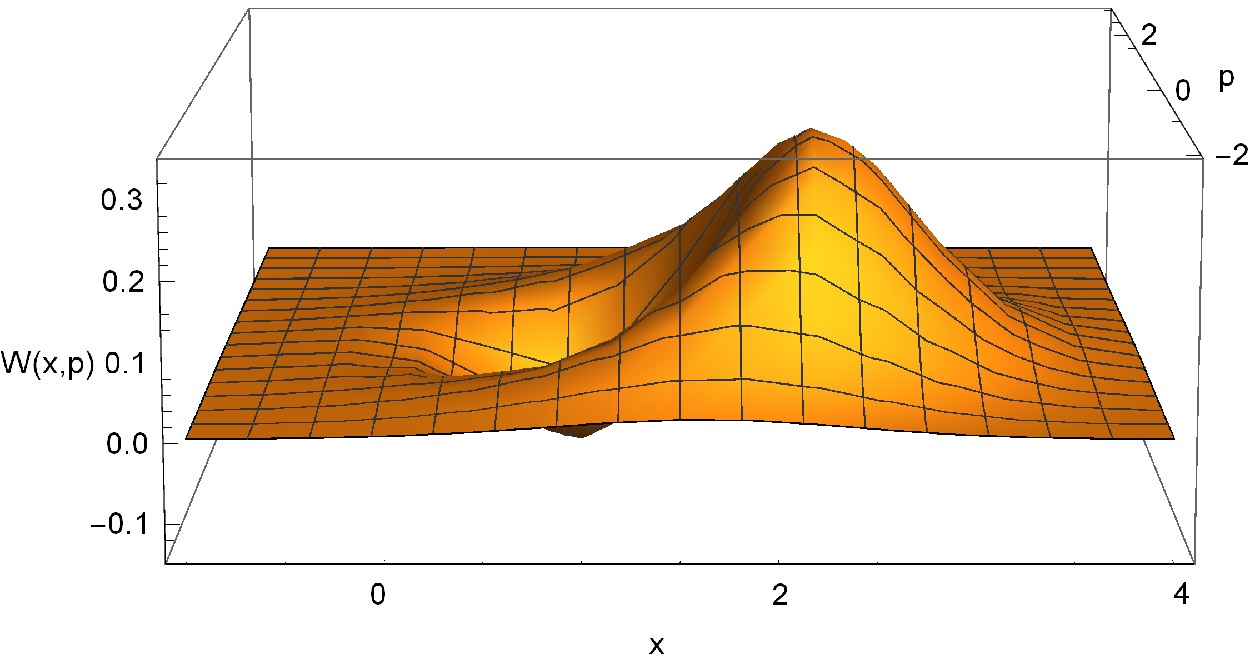}}
\end{subfigure}
\caption{Wigner functions for a standard coherent state (top) and a photon added coherent state (bottom) for values $|\alpha|^2=1$ and $T=1$.}
\label{fig:Wigs}
\end{figure}

A similar argument can be made for photon addition. Shown in Fig. \ref{fig:paddcoh}, multiple photon additions shift the number distribution of a coherent state as more photons are added. While the effect of adding a single photon on a coherent state's average photon number is not as drastic as that for other states, the result is still somewhat unintuitive as the average photon number for a single photon added coherent state is $\langle \hat{n}_{1+} \rangle=T|\alpha|^2+2-\frac{1}{1+T|\alpha|^2}$, where $|\alpha|^2$ is the average photon number in the initial coherent state. In general it appears as the average photon number increases by almost two, even when only a single photon is added to the beam. This trend extends to higher photon added coherent states and has a form of 
\begin{equation}
\langle \hat{n}_{m+} \rangle=T|\alpha|^2+2m-\frac{m L_{m-1}(-T|\alpha|^2)}{L_m(-T|\alpha|^2)},
\label{eq:ncoh+}
\end{equation}
where $m$ is the number of added photons.
Since we are working with Wigner functions, it is instructive to compare the change undergone in the Wigner function, even under only a single photon addition. This effect can be seen in Fig. \ref{fig:Wigs}. We note that the Wigner function for a photon added coherent state, attains negative quasi-probability values, a sufficient, but not necessary, condition of a non-classical quantum state.

Before proceeding to discuss the performance of various metrics, it is first important to mention the behavior of these post selection events and how their statistics behave in regards to the probability of successfully generating these states. We note that Fig. \ref{fig:paddcoh} and \ref{fig:Wigs} have shown an ideal case, setting the transmissivity of the variable beam splitter to $T=1$. In this treatment, generally the resulting state under this condition has the ``best" characteristics. However, its clear that generating a photon added or subtracted state under these conditions is extremely unlikely, as, if the beam splitter transmissivity is set to unity, no photon addition or subtraction event is likely to occur. This is characterized and shown in Fig. \ref{fig:probscoh}. One can also see this from the form of Eq. \ref{eq:paddm} that $P_m \rightarrow 0$ for $T \rightarrow 1$. It is also in this limit that this beam splitter model for both photon addition and subtraction converge back to the mathematical treatment of the creation and annihilation operators (for PDC addition, this connection is made when $G\rightarrow 1$). For values of $T<1$ ($G>1$) we then create a photon added or subtracted state with some less than ideal statistics. Of course the resulting state itself also is modified as $T$ ($G$) changes and, in general, the characteristics of the state worsen as $T$ ($G$) approaches to zero (infinity). However, the question remains, is there a region where the probability of creating some photon added or subtracted state remains significantly large and the resulting state also contains some useful character? In order to answer this question, we investigate various metrics. Many different metrics may be used when characterizing a quantum metrology scheme. Here we will discuss phase information (phase sensitivity) through Fisher Information as well as Signal to Noise Ratio. Generally the phase estimation route is viewed as more robust and is typically the chosen metric in quantum metrology, but we will see that SNR, while being perhaps a more limited metric, has some aspects not possible when phase estimation is considered. In practice, phase information is useful for applications of measuring an unknown parameter inside the interferometer, where SNR is more applicable to imaging or ranging. Here we will focus mostly on the SNR, as phase information has strict bounds that, for the previously specified topology, cannot beat the standard classical limit.

As a general improvement, even a probabilistic amplification is bounded by the Quantum Cram\'{e}r-Rao Bound (QCRB), which, for a fixed MZI topology, is calculated as a function of the input states only, and minimizes (maximizes for the Quantum Fisher Information (QFI)) over all possible detection schemes \cite{Caves1994}. Here, since our only modification of the MZI is after the unknown phase, $\phi$, it can be viewed as a particular measurement scheme, and so the QCRB is most directly calculated as a function of the total state, immediately after the unknown phase $\phi$, depicted in Fig. \ref{fig:MZI}. When this is calculated for any classical states as input states, the full state is Gaussian and thus can be calculated following \cite{QFI1}. As expected, since the states at this point are purely classical, the QCRB is simply the Shot-Noise Limit (SNL), which is given by $I_Q^{-1}=\frac{1}{\nu \bar{n}}=\textrm{QCRB}=\Delta \phi_{min}^2$, where $\bar{n}$ is the average photon number in the initial state and $\nu$ is the number of experimental trials. This result seems discouraging, as it shows that one cannot improve overall phase variance with this probabilistic amplification. However, since this probabilistic amplification is a form of weak value amplification and it is known that weak value amplification schemes show their benefit when technical noise is considered, one may still be able to use this implementation under the conditions of certain technical noise \cite{Howell2014}, but will not be considered here. For the reasons discussed here, we will now focus on the SNR as our chosen metric.

\begin{figure}[!htb]
\includegraphics[width=\columnwidth]{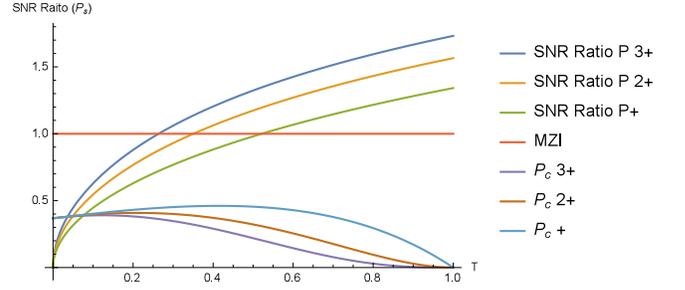}
\caption{SNR ratios of photon added coherent states to a normal coherent state are shown for $|\alpha|^2=1$. Also shown are the probabilities of successfully generating these states with the beam splitter implementation. SNR ratios above one show a superior SNR when compared to a normal coherent state, which exist for significantly high probability.} \label{fig:SNRcoh}
\end{figure}

The SNR, defined by $\textrm{SNR}=\frac{\langle \hat{a}^\dagger\hat{a} \rangle}{\sigma_{\hat{a}^\dagger\hat{a}}}$, that is the average photon number divided by its standard deviation, also typically described in terms of intensity. Since this metric is not constructed in the same way as Fisher Information, repeated trials and asymptotic limits are not necessarily incorporated into this metric. Therefore, while we should still note the associated probabilistic nature of the processes here, they are not directly incorporated into the metric. In the case of photon added states, the SNR is defined to be  $\textrm{SNR}_{m+}=\frac{\langle \hat{a}^\dagger\hat{a} \rangle - m}{\sigma_{\hat{a}^\dagger\hat{a}}}$, where the number of added photons, $m$, is subtracted out of the signal to only consider the signal of the resulting photon added state itself. In the case of a photon added coherent state, as discussed earlier, the average photon number increases by approximately $2m$ and therefore the resulting signal is still enhanced by $m$, which is significant in the low photon regime. In order to get a general expression for the SNR of a $m$ photon added coherent state, along with Eq. \ref{eq:ncoh+}, we only need the general expression for the second moment for a $m$ photon added coherent state. While noticeably more complicated than the first moment, it is found to be,
\begin{equation}
\begin{split}
\langle \hat{n}_{m+}^2 \rangle=&\frac{1}{L_m(-T|\alpha|^2)}\times \\
&( (m+2)(m+1)L_{m+2}(-T|\alpha|^2)\\
&-3(m+1)L_{m+1}(-T|\alpha|^2)+L_m(-T|\alpha|^2) )
\end{split}
\end{equation}
From this equation and Eq. \ref{eq:ncoh+} we then can easily construct a general form for the SNR of a $m$ photon added coherent state, which takes the form
\begin{equation}
\textrm{SNR}_{m+}=\frac{\langle \hat{n}_{m+} \rangle-m}{\sqrt{\langle \hat{n}_{m+}^2 \rangle-\langle \hat{n}_{m+} \rangle^2}}
\end{equation}
Consider Fig. \ref{fig:SNRcoh}, where we have formed a ratio of the SNR of various photon added coherent states to the SNR of a normal coherent state ($\textrm{SNR}_\alpha=|\alpha|$). As shown, there is a region where the SNR ratios attain a superior SNR (above 1) while maintaining a respectable probability of success.

This analysis thus shows some promise for the scheme, when SNR is chosen as the metric.

\begin{figure}[!htb]
\includegraphics[width=\columnwidth]{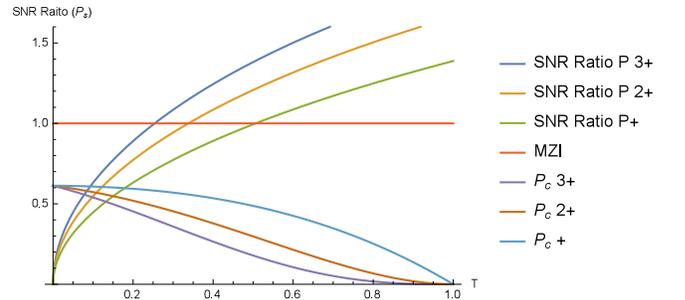}
\caption{SNR ratios of photon added coherent states to a normal coherent state are shown for $|\alpha|^2=1$, photon loss $L=0.3$ and detector efficiency $D=0.7$. Also shown are the probabilities of successfully generating these states with the beam splitter implementation. SNR ratios above one show a superior SNR when compared to a normal coherent state, which exist for significantly high probability.} \label{fig:loss}
\end{figure}

A look at how this scheme performs under loss shows, at first glance, a fairly surprising result, shown in Fig. \ref{fig:loss}. Comparing this figure to the ideal case in Fig. \ref{fig:SNRcoh}, it seems surprising that this scheme performs better, relative to a standard coherent state, under lossy conditions. The explanation here is that under a simple model of loss, assuming all photon loss occurs equally in both arms of the MZI and that all detectors share the same efficiency, these assumptions simply amount to a change of variables of the form $\alpha \rightarrow (1-L)D\alpha$, where $0 \leq L \leq 1$ is the photon loss and $0 \leq D \leq 1$ is the detector efficiency. This model of loss then simply works to lower the average photon number arriving at the detectors, but we know that this scheme actually performs best at low photon number as the effect of photon addition is most significant in this regime and thus, our SNR ratio actually improves in this case.

\begin{figure}[!htb]
\includegraphics[width=\columnwidth]{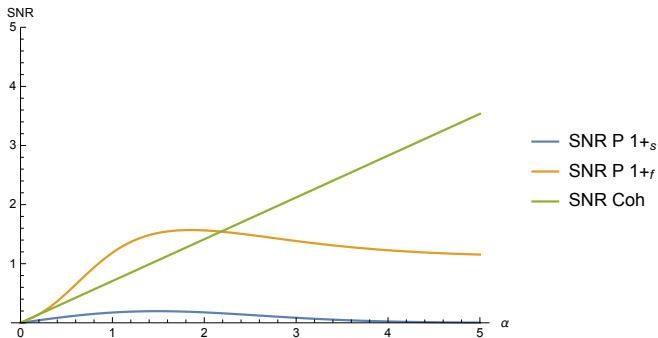}
\caption{Plot of weighted $\sqrt{P_i}\times \textrm{SNR}_i$ for both possible outcomes of attempted photon addition via our variable beam splitter model. Each SNR is weighted by its square root probability of being generated. Parameters have been set to $T=0.05$ and $\phi=\pi/2$. Surprisingly, it is the failed attempts at addition that have superior SNR, which implies that this metric may not capture the full implications of post selection.} \label{fig:comp1}
\end{figure}

\begin{figure}[!htb]
\includegraphics[width=\columnwidth]{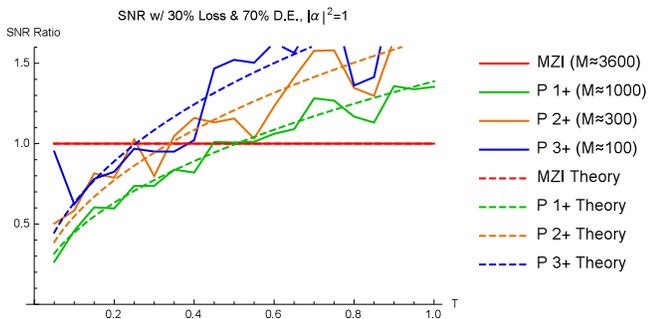}
\caption{Plot of SNR ratio, similar for Fig. \ref{fig:loss}, but with simulated data for a number of measurements of 3600 per value of tranmissivity and the associated successful measurements (M) for each scheme. For each value of tranmissivity, we have averaged over all successful measurements.} \label{fig:data1}
\end{figure}

It has been suggested that a metric of $\sqrt{P_s} \times \textrm{SNR}_s$ be used for post selection when SNR is the chosen metric  \cite{Pandey2013}. However, if one considers all outcomes in this way, while $\sqrt{P_s} \times \textrm{SNR}_s$ performs quite poorly, the complementary case, $\sqrt{P_f} \times \textrm{SNR}_f$ actually performs significantly better, as shown in Fig. \ref{fig:comp1}. This result is surprising and may indicate that this choice of metric is not necessarily the best choice. This leads us to a more simulated data driven metric as shown in Fig. \ref{fig:data1}.

While the previous figures show promise for the SNR ratio as a function of the transmissivity, along with the associated probabilities for each measurement condition, it is perhaps more useful to look at simulated data, to showcase how these schemes, along with their inherent probabilities are expected to perform in an experiment. Fig. \ref{fig:data1} shows simulated data for the same cases as the above (same loss and efficiency parameters), along with dashed curves showing the theory representations.  The total number of measurements is fixed at $3600$ measurements for all the displayed states, with $M$ showing the number of ``kept measurements" for each case, due to the post selection requirement  (this is simply related by each of the $P_m$'s given by Eq. \ref{eq:paddm}). After averaging over all kept measurements for each value of tranmissivity, we see that the higher photon additions do attain a better SNR as predicted by theory, but the ``scatter" also worsens for the higher photon additions as the kept measurements also decreases significantly.

In this discussion we have seen that the use of post selection and nondeterministic, noiseless amplification, at the output of an MZI, shows some promise when considering a coherent state as input and SNR as a metric. We have also shown that if the metric of phase information is used, a standard in quantum metrology, then the probabilistic nature of post selection must be taken into account when claims of improvement are made. This consideration agrees with many other claims that post selection alone, cannot increase total information and must necessarily discard some information, limiting this scheme to the QCRB, which shows a classical limit. However, turning to a discussion of SNR and perhaps applications in imaging or ranging, this method shows an improvement over standard techniques. We have also shown that this scheme does not rely on ideal conditions and actually performs better under lossy conditions as compared to a normal coherent state. We have also shown that, within some scatter tolerance, simulated data shows agreement with that of theory and illustrates the effect of the non-deterministic nature of this scheme.
\\

%
% Acknowledgments
%

\begin{acknowledgments}
The research in this document is from the Quantum Metrology Innovations for Military Missions program,
which is being developed by the authors with funding from the Defense Advanced Research Projects Agency (DARPA). The views, opinions, and/or findings contained in this presentation are those of the authors and should not be interpreted as representing the official views or policies of the Department of Defense or the U.S.Government. BTG would like to acknowledge support from the National Physical Science Consortium \& National Institute of Standards and Technology graduate fellowship program, the Boeing corporation, as well as helpful discussions with Dr. Emanuel Knill at NIST-Boulder. J.P.D would like to acknowledge support from the Air Force Office of Scientific Research, the Army Research Office, the Boeing Corporation, the National Science Foundation, and the Northrop Grumman Corporation. 
\end{acknowledgments}

%
% Bibliography
%
\bibliography{bib}

\end{document}